\documentclass[prl,aps,twocolumn,10pt,showpacs,groupedaddress,superscriptaddress,floatfix]{revtex4-1}
\usepackage{amsmath,amsfonts,amssymb,graphics,graphicx,epsfig,color,times}
\usepackage[utf8x]{inputenc}
\usepackage{color}
\usepackage{bbm, dsfont} 
\usepackage{subfigure}
\usepackage{hyperref}
\usepackage{mathrsfs}
\usepackage{verbatim}
\begin{document}
\title{Superradiant laser: Effect of long-ranged dipole-dipole interaction}
\author{H. H. Jen}
\email{sappyjen@gmail.com}
\affiliation{Institute of Physics, Academia Sinica, Taipei 11529, Taiwan}
\date{\today}
\renewcommand{\r}{\mathbf{r}}
\newcommand{\f}{\mathbf{f}}
\renewcommand{\k}{\mathbf{k}}
\renewcommand{\r}{\mathbf{r}}
\def\bea{\begin{eqnarray}}
\def\eea{\end{eqnarray}}
\def\ba{\begin{array}}
\def\ea{\end{array}}
\def\bdm{\begin{displaymath}}
\def\edm{\end{displaymath}}
\def\red{\color{red}}
\begin{abstract}
We theoretically investigate the effect of long-ranged dipole-dipole interaction (LRDDI) on a superradiant laser (SL).\ This effect is induced from the atom-photon interaction in the dissipation process.\ In the bad-cavity limit usually performed to initiate SL, we demonstrate that cavity photon number oscillates as an inter-particle distance of the atoms varies.\ Similarly the atom-atom coherence alternates with signs, showing critical transitions alternatively in SL operations.\ This suggests a complexity of the collective effect emerging in a large ensemble of atoms.\ Therefore this effect in a SL can not be simply interpreted by only a part of the whole ensemble.\ We numerically solve for a steady state SL including the spatially-dependent LRDDI, and locate the maximal cavity photon number and the minimal spectral linewidth respectively at the optimal atomic separations in the setting of an equidistant atomic array.\ The scaling of a finite number of atoms shows an outperformed steady state SL than the one without LRDDI, which allows for probing narrow atomic transitions and is potentially useful for precision measurements and next generation optical clocks.
\end{abstract}
\pacs{42.50.Nn, 42.50.Ct, 37.30.+i, 42.50.Gy}
\maketitle
{\it Introduction.} 
Superradiant laser (SL) \cite{Haake1993} allows for high spectral purity and robust frequency standard, which is crucial for precision measurements, tests of fundamental physics, and next generation optical clocks \cite{Ludlow2015}.\ In contrast to a conventional laser in the good cavity limit, steady-state SL \cite{Meiser2009, Meiser2010, Bohnet2012} in a high-Q cavity is predicted to have an extremely narrow linewidth \cite{Meiser2009}.\ This linewidth exceeds the limitation set by thermal fluctuations of cavity mirrors due to the dominance of atomic coherences in the bad cavity limit \cite{Kuppens1994, Chen2009}.\ From these collective atomic dipoles in such driven atom-cavity system, SL enables a continuous and collective emission with a high degree of phase coherence.\ This synchronization of dipoles inside a cavity thus operates in a new quantum regime of laser operations, where even less than one intracavity photon is mediating the gain medium \cite{Meiser2009, Bohnet2012}. 

The build-up of atomic coherences is initiated from the collective light-matter interactions which can lead to superradiance \cite{Dicke1954, Gross1982, Mandel1995} that is attributed to the effect of induced long-ranged dipole-dipole interaction (LRDDI) \cite{Stephen1964, Lehmberg1970}.\ Recently a redshift in the cooperative spontaneous emission is observed due to this collective light-matter interaction in various atomic systems including the planar cavity \cite{Rohlsberger2010}, an atomic vapor \cite{Keaveney2012}, an ionic atomic array \cite{Meir2014}, and a cold atomic ensemble \cite{Pellegrino2014}.\ These measurements agree well with theoretical predictions, and signify the essence of induced dipole-dipole interaction either in the mean-field \cite{Keaveney2012} or many-body regimes \cite{Meir2014, Pellegrino2014}.\ Aside from the aforementioned enhanced spontaneous emission or its associated collective frequency shift \cite{Friedberg1973, Scully2009, Jen2015}, a subradiant emission can also be observed in a large cloud of atoms \cite{Guerin2016} as a afterglow \cite{Mazets2007} of superradiance.\ Recent proposals to prepare such collective single-photon subradiant states \cite{Eberly2006, Mazets2007, Svidzinsky2008} involve an ultrafast control of the phases on the atoms \cite{Scully2015} and a coherent manipulation of the phase imprinting in a one-dimensional atomic array \cite{Jen2016_SR}.\ 

What is more intriguing in this dynamically and collectively coupled light-matter interacting system is the role of atom-atom correlations in a large atomic ensemble \cite{Jennewein2016}.\ The crossover from negligible to emerging atomic coherence marks the onset of significant atom-atom interactions that mean-field approximation of a polarizable medium fails \cite{Jenkins2016}.\ Here we investigate this effect from the induced LRDDI on the steady state SL, and demonstrate the influence of atom-atom correlations on the performance of the SL.\ This many-body correlation induced by collective decay rates and frequency shifts is often regarded as negligible \cite{Meiser2009} or is simply parametrized by one characteristic decay rate \cite{Meiser2010}.\ This collective effect has been studied for a SL in an optical lattice \cite{Maier2014}, which tends to broaden the spectral linewidth away from deep in the bad-cavity limit.\ We include this LRDDI, as is ubiquitous in atom-photon interactions, to generalize the mechanism of cavity-QED system that enables SL.\ This generalization should allow us to investigate a less explored regime of SL, and help to locate the optimal operation parameters for its best performance. 

\begin{figure}[t]
\centering
\includegraphics[width=8.0cm,height=5.5cm]{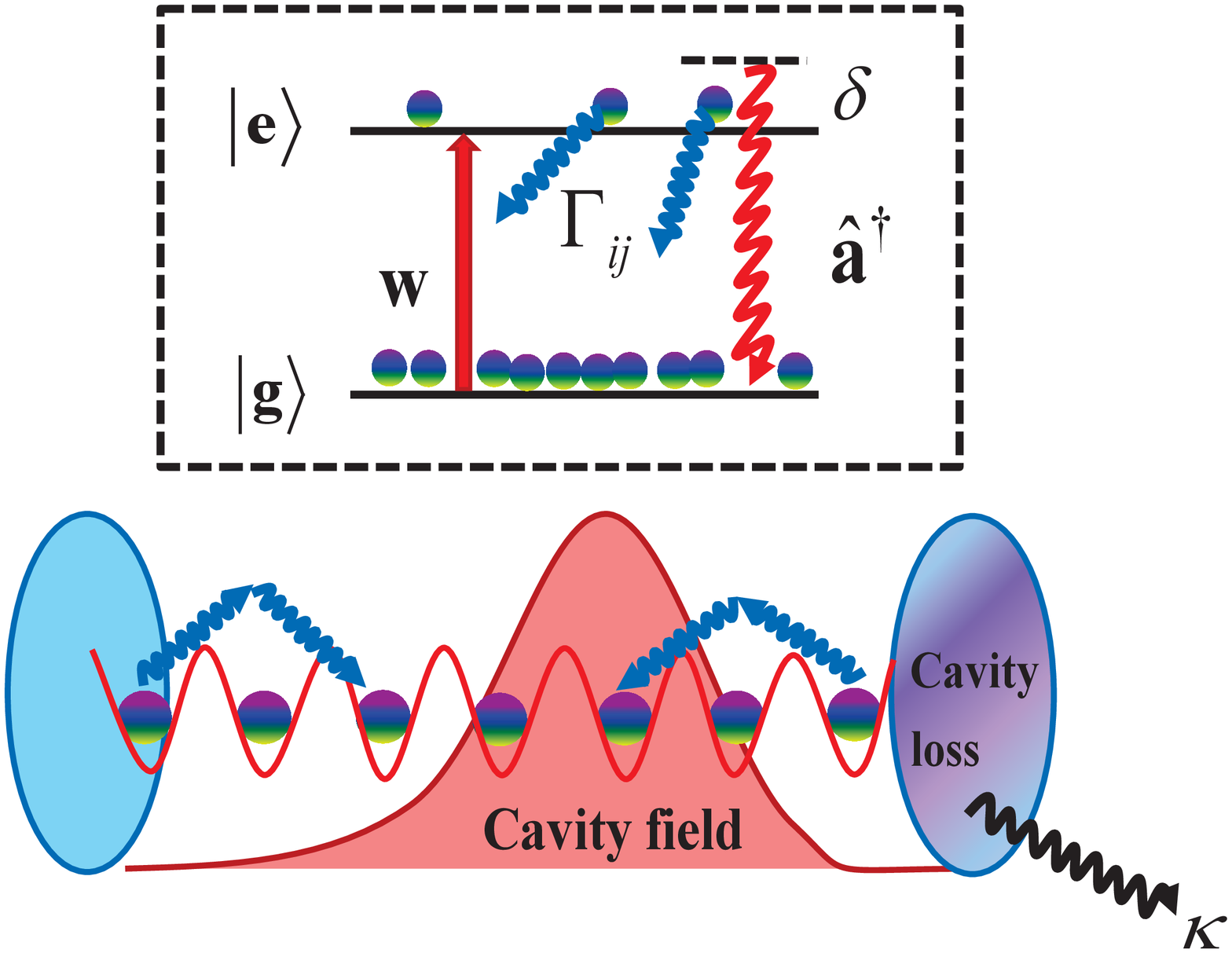}
\caption{(Color online) Schematic cavity-QED system with long-ranged dipole-dipole interaction.\ The incoherent repumping field ($w$) drives the atoms to the excited states $|e\rangle$ from the ground states $|g\rangle$.\ Cavity photon is denoted as a bosonic operator $\hat{a}$ with a loss rate $\kappa$ and has a detuning $\delta$ from the transition frequency.\ Cooperative spontaneous emission $\Gamma_{ij}$ is long-ranged in nature induced from the resonant dipole-dipole interaction.}\label{fig1}
\end{figure}

{\it Hamiltonian for SL.}
The theoretical analysis is based on the Hamiltonian ($H$) of a cavity-QED system and the Lindblad forms of the decay and repumping processes for effective two-level atoms ($|g\rangle$ and $|e\rangle$ for the ground and excited states respectively).\ Here we consider a general form of the decay process that involves the induced dipole-dipole interaction \cite{Lehmberg1970, Jen2016_SR} as shown in Fig. \ref{fig1}.\ This LRDDI originates from the rescattering events in the common quantized field, which results in a pairwise collective frequency shift $G_{\mu\nu}$ and oscillatory decay rate $F_{\mu\nu}$ with a dependence of an atomic distance $|\r_\mu-\r_\nu|$.\ The Hamiltonian of a cavity-QED system along with a collective frequency shift reads
\bea
H&=&\omega_a\sum_{\mu=1}^N \hat{\sigma}^{ee}_\mu+\omega_c \hat{a}^\dag\hat{a}+\frac{g}{2}\sum_{\mu=1}^N(\hat{a}^\dag\hat{\sigma}^-_\mu+\hat{\sigma}^+_\mu \hat{a})\nonumber\\
&+&\sum_{\mu\neq\nu}^N\sum_{\nu=1}^N G_{\mu\nu}\hat{\sigma}_\mu^+\hat{\sigma}_\nu^-,
\eea
where we let $\hbar$ $=$ $1$, and the lowering (raising) operator is $\hat{\sigma}_\mu^-$ ($\hat{\sigma}_\mu^+$) where $\hat{\sigma}_\mu^-$ $\equiv$ $|g\rangle_\mu\langle e|$ and $\hat{\sigma}_\mu^-$ $\equiv$ $(\hat{\sigma}_\mu^+)^\dag$.\ Excited state population operator is $\hat{\sigma}_\mu^{ee}$ with a transition frequency $\omega_{a}$.\ The atom-cavity coupling constant is $g$ $=$ $d\mathcal{E}(\omega_c)$ where $d$ is the dipole moment, and $\mathcal{E}(\omega_c)$ $\equiv$ $\sqrt{\omega_c/(2\epsilon_0 V)}$ with a quantization volume $V$.\ Single mode cavity field is $\hat{a}$, and a complete expression of $G_{\mu\nu}$ can be found below.

We consider the cooperative decay process that is long-ranged in nature and an incoherent repumping field in the Heisenberg equations for arbitrary quantum operators $\hat{Q}$, 
\bea
\frac{d\hat{Q}}{dt} = \frac{1}{i\hbar}[\hat{Q},H] + \left(\mathcal{L}_r + \mathcal{L}_c + \mathcal{L}_s\right)[\hat{Q}], 
\eea
where various Lindblad forms for a repumping (r) field, a cavity (c) loss, and spontaneous (s) emission processes are 
\bea
\mathcal{L}_r[\hat{Q}]&=& -\frac{w}{2}\sum_{\mu=1}^N\left(\hat{\sigma}_\mu^-\hat{\sigma}_\mu^+\hat{Q}+\hat{Q}\hat{\sigma}_\mu^-\hat{\sigma}_\mu^+-2\hat{\sigma}_\mu^-\hat{Q}\hat{\sigma}_\mu^+\right),\\
\mathcal{L}_c[\hat{Q}]&=&-\frac{\kappa}{2}\left(\hat{a}^\dag\hat{a}\hat{Q}+\hat{Q}\hat{a}^\dag\hat{a}-2\hat{a}^\dag\hat{Q}\hat{a}\right),\\
\mathcal{L}_s[\hat{Q}]&=&-\sum_{\mu,\nu=1}^N\frac{F_{\mu\nu}}{2}\left(\hat{\sigma}_\mu^+\hat{\sigma}_\nu^-\hat{Q}+\hat{Q}\hat{\sigma}_\mu^+\hat{\sigma}_\nu^- -2\hat{\sigma}_\mu^+\hat{Q}\hat{\sigma}_\nu^-\right).\nonumber\\
\eea
Here a repumping and a cavity loss rates are $w$ and $\kappa$ respectively, and $F_{\mu\nu}$, $G_{\mu\nu}$ are expressed as \cite{Lehmberg1970}
\bea
F_{\mu\nu}(\xi)&\equiv&
\frac{3\Gamma}{2}\bigg\{\left[1-(\hat{d}\cdot\hat{r}_{\mu\nu})^2\right]\frac{\sin\xi}{\xi}\nonumber\\
&+&\left[1-3(\hat{d}\cdot\hat{r}_{\mu\nu})^2\right]\left(\frac{\cos\xi}{\xi^2}-\frac{\sin\xi}{\xi^3}\right)\bigg\},\label{F}\\
G_{\mu\nu}(\xi)&\equiv&\frac{3\Gamma}{4}\bigg\{-\Big[1-(\hat{d}\cdot\hat{r}_{\mu\nu})^2\Big]\frac{\cos\xi}{\xi}\nonumber\\
&+&\Big[1-3(\hat{d}\cdot\hat{r}_{\mu\nu})^2\Big]
\left(\frac{\sin\xi}{\xi^2}+\frac{\cos\xi}{\xi^3}\right)\bigg\}\label{G}, 
\eea
where $\Gamma$ is the effective spontaneous decay rate, $\xi$ $=$ $|\mathbf{k}| r_{\mu\nu}$, and $r_{\mu\nu}$ $=$ $|\mathbf{r}_\mu-\mathbf{r}_\nu|$ with the transition wave vector $|\mathbf{k}|$.\ The self-consistent coupled Heisenberg equations can be found in Supplemental Material (SM) \cite{SM}, which we numerically solve for the steady state solutions.\ 

For $N$ number of atoms, there are $2N$ $+$ $C^N_2$ $+$ $1$ variables ($C$ denotes the binomial coefficients) in the Heisenberg equations, which involve $N$ individual atomic excited state populations and atom-photon coherence operators respectively, $C^N_2$ atom-atom coherences, and one cavity field.\ This invokes generally a non-polynomial dependence of number of coupled equations ($\sim$ $N^2$ for a large $N$), which points to the complexity of "More is different" \cite{Anderson1972} similarly here in the atom-cavity systems.\ To get around the complexity from intractable many-body coherences induced from an atom-photon interaction, we take a bottom-up approach by adding one atom at a time into the atomic system to investigate the effect of LRDDI on a SL.

\begin{figure}[t]
\centering
\includegraphics[width=8.0cm,height=4.5cm]{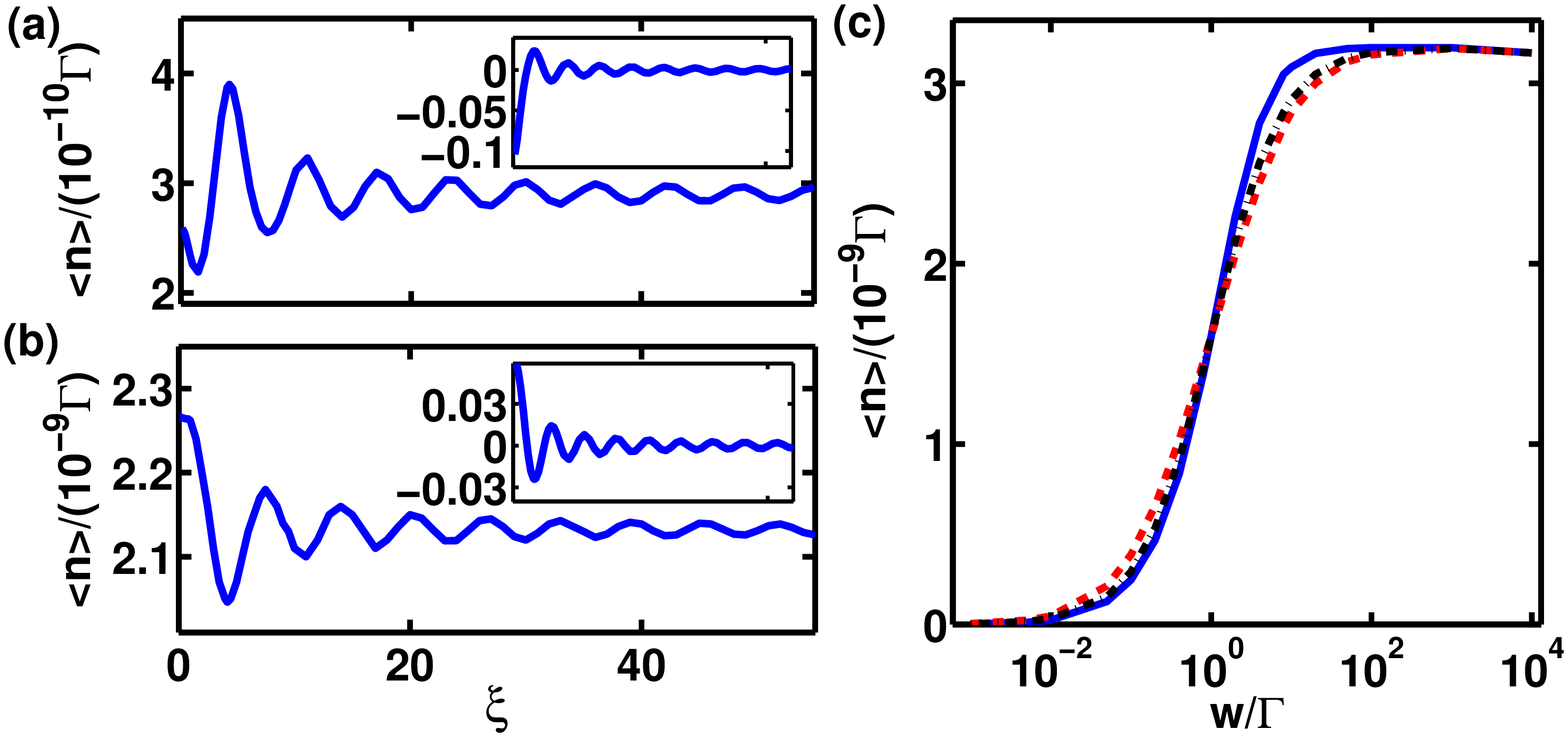}
\caption{(Color online) Cavity field for SL of $N$ $=$ $2$.\ Deep in the bad-cavity limit we choose the coupling constant $g$ $=$ $40\Gamma$ with a zero detuning $\delta$, and cavity loss rate $\kappa$ $=$ $10^6\Gamma$.\ The cavity photon number $\langle n\rangle$ $\equiv$ $\langle \hat a^\dag \hat a\rangle$ has an oscillatory dependence of atomic separation $\xi$ for repumping fields (a) $w$ $=$ $0.1$ and (b) $2\Gamma$.\ The insets are atomic coherences, showing oscillatory sign changes as $\xi$ varies.\ (c) The dependence of photon number of $w$ in logarithmic scales for $\xi$ $=$ $0.5$ (solid), $4$ (dash), and $100\lambda$ (dash-dots).}\label{fig2}
\end{figure}

{\it Cavity photons and linewidth.} In the mean-field treatment \cite{Meiser2009} where LRDDI is negligible, the increase of cavity fields indicates a transition that SL is enabled along with the emergence of sign change in the atom-atom coherence.\ Also the number of photons inside a cavity in the bad-cavity limit should be much less than one.\ To investigate the effect of LRDDI, first we consider a SL of two atoms aligning along a cavity axis and choose a perpendicular dipole orientation, that is $\hat d\cdot\hat r_{\mu\nu}$ $=$ $0$, without loss of generality.\ In Fig. \ref{fig2}(a) and (b), we plot the cavity photon number $\langle n\rangle$ as the inter-particle distance ($\sim\xi$) varies.\ The oscillatory feature is expected due to the sinusoidal dependences in collective decay rate and frequency shift of Eqs. (\ref{F}) and (\ref{G}).\ Less than one intracavity photon ($10^{-9}\Gamma/\omega_a$ $\ll$ $1$) is also predicted in the bad-cavity limit when $\kappa$ $\gg$ $g$ and $w$.\ The photon number approaches a constant value, reaching the noninteracting regime when LRDDI is less significant at large $\xi$.\ 

At small inter-particle separation, the maximal photon number emerges at $\xi$ $\approx$ $4$ for a weak repumping rate as shown in Fig \ref{fig2}(a).\ In contrast for a large repumping rate in Fig. \ref{fig2}(b), the cavity field maximizes as $\xi$ $\rightarrow$ $0$.\ The insets are real parts of the atomic coherence operator $\langle\hat{\sigma}_{1}^+\hat\sigma^-_2\rangle_c$ in steady states, where $c$ denotes the cumulant of correlation defined in SM \cite{SM}.\ The sign changes of the atomic coherence, positive to negative and vice versa alternatively, reflect the corresponding increase or decrease of photon fields.\ This alternating atomic coherence also suggests of complexity of the collective effect in a large ensemble of atoms where atoms distribute randomly.\ Therefore the collective effect in SL can not be simply interpreted by only a part of the whole ensemble.\ In Fig. \ref{fig2}(c) the cavity field is enhanced as the repumping rate increases, which saturates at $w$ $\gtrsim$ $100$.\ The photon number dominates respectively at finite and small atomic separations for weak and strong repumping rates.\ The regime separating weak and strong $w$ can be seen at the crossing point of $w$ $=$ $1\Gamma$, showing the same critical transition of an enhanced photon number in the mean-field treatment of a large ensemble of atoms \cite{Meiser2009}.\ The reason why the critical condition occurs at $w$ $=$ $1\Gamma$ is due to the disappearance of LRDDI in a SL (see coupled Heisenberg equations in SM \cite{SM}) when all atoms in steady states are half excited ($\langle\hat\sigma^{ee}_\mu\rangle_c$ $\approx$ $1/2$) and atom-atom coherences are negligible ($\langle\hat{\sigma}_{1}^+\hat\sigma^-_2\rangle_c$ $\approx$ $10^{-6}$).\ We note that the oscillatory patterns are similar respectively in the weak ($w$ $\lesssim$ $1\Gamma$) and strong regimes ($w$ $\gtrsim$ $1\Gamma$) of repumping rates except at around the critical $w$ where oscillation diminishes.\ Also for more atoms up to $N$ $=$ $5$, we find similar characteristics of cavity fields generated from an equidistant atomic array \cite{SM}.\ 

Next we calculate the spectral linewidth of the photon field from its first-order correlation which leads to the power spectrum $S(\nu)$ $=$ $\pi^{-1}$ Re$[\int_0^\infty d\tau\langle\hat a^\dag(0)\hat a(\tau)\rangle e^{i\nu\tau}]$ \cite{QO:Scully}.\ The time-evolved correlation can be derived by quantum regression theorem \cite{QO:Scully, Tannoudji1992}.\ Define $A(t)$ $\equiv$ $\langle\hat{a}^\dag(t)\hat{a}(0)\rangle$ and $B_\mu(t)$ $=$ $\langle\hat{\sigma}_\mu^+(t)\hat{a}(0)\rangle$, we have 
\bea
\frac{dA(t)}{dt}&=&(i\delta-\frac{\kappa}{2})A(t)+\frac{ig}{2}\sum_{\mu=1}^NB_\mu(t),\nonumber\\
\frac{dB_\mu(t)}{dt}&=&-\frac{w+\Gamma}{2}B_\mu(t)-\frac{ig}{2}(2\hat{\sigma}_{\mu,s}^{ee}-1)A(t)\nonumber\\
&-&\sum_{\nu\neq\mu}^N\frac{F_{\mu\nu}-i2G_{\mu\nu}}{2}(1-2\hat{\sigma}_{\mu,s}^{ee})B_\nu(t),
\eea
where $A(0)$, $B_\mu(0)$, and $\hat{\sigma}_{\mu,s}^{ee}$ are initial conditions, which can be derived from the steady state solutions of Heisenberg equations.\ We then determine the spectral linewidth from the FWHM (full-width at half-maximum) of $S(\nu)$.

\begin{figure}[t]
\centering
\includegraphics[width=8.0cm,height=4.5cm]{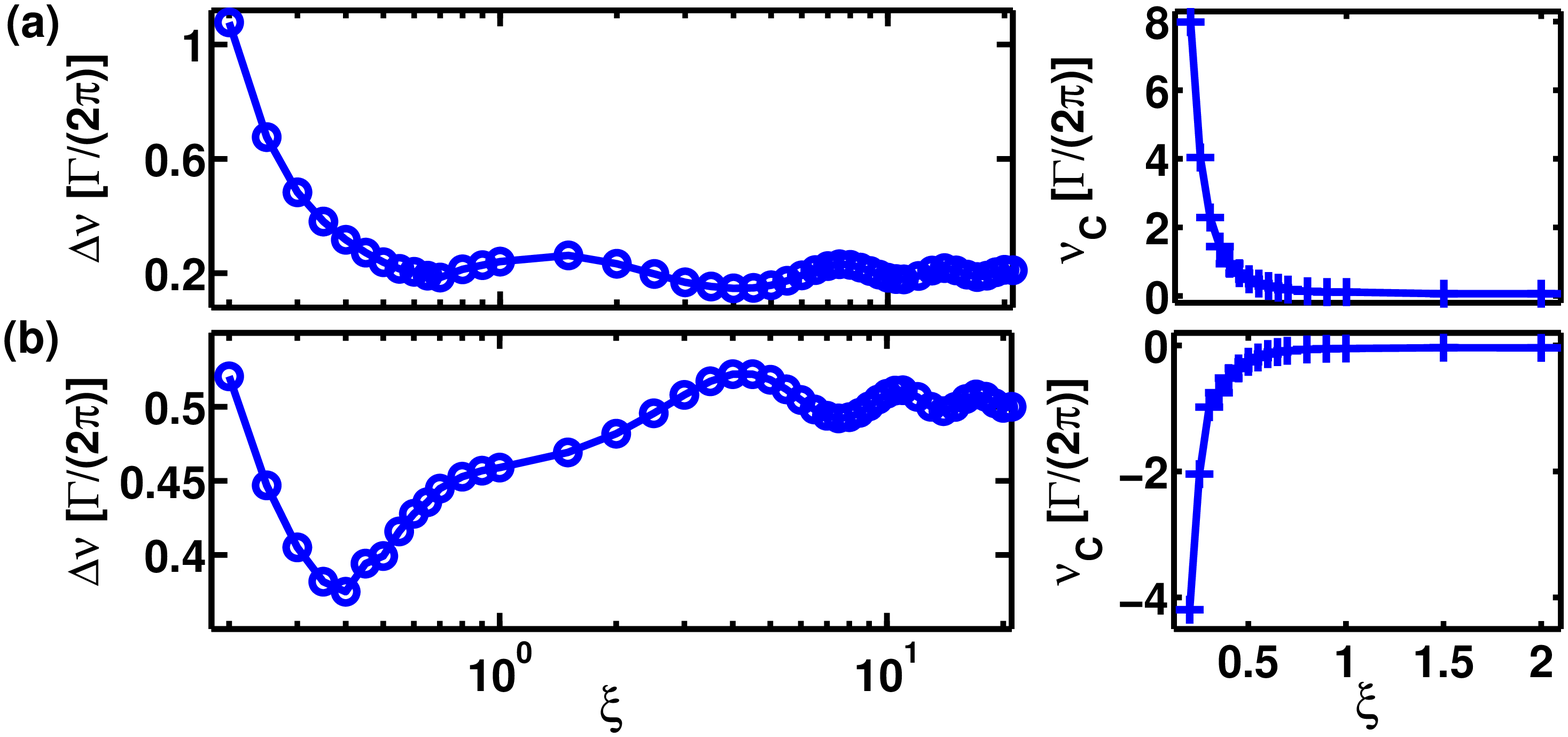}
\caption{(Color online) Spectral linewidth ($\Delta\nu$) and central frequency shift ($\nu_C$) for a SL of $N$ $=$ $2$.\ The linewidths decrease and reach the minimum at some finite atomic separations $\xi$'s, and approach asymptotically the noninteracting regime where $\xi$ $\gg$ $10\lambda$ for (a) $w$ $=$ $0.1$ and (b) $2\Gamma$.\ Corresponding central frequency shifts of cavity fields are plotted in the right panels, indicating blue- and red-shifts respectively.}\label{fig3}
\end{figure}

The linewidth ($\Delta\nu$) and associated frequency shift ($\nu_C$) of the cavity photons are shown in Fig. \ref{fig3}.\ The minimum of linewidths for a weak repumping rate appears at $\xi$ $\approx$ $4$, corresponding to the maximum of $\langle n\rangle$.\ For a strong $w$, the minimum in contrast occurs at a smaller $\xi$ $\approx$ $0.4$.\ Comparing the minimal linewidths with the ones in a noninteracting regime at large $\xi$, for both weak and strong $w$, they have a reduction of $\sim 30\%$ due to LRDDI at finite $\xi$'s.\ Moreover in Fig. \ref{fig3}(b), the linewidths in a region of $\xi$ $\approx$ $0.3-2$ outperform the one in a noninteracting regime.\ This demonstrats a broad regime of operation parameters that allow a SL with narrow linewdiths.

For even smaller inter-particle distances where LRDDI is more significant, the spectral linewidths increase and exceed the noninteracting values.\ The corresponding frequency shifts also show a stronger dependence at $\xi$ $\lesssim$ $1$.\ The effect of LRDDI manifests on the reduced linewidth at the optimal inter-particle distance, which is superior to the noninteracting regime and is favorable for probing ultranarrow atomic transitions.\ For more number of particles $N$ in the setting of an equidistant atomic array, we find again they have a similar dependence on $\xi$ as in Fig. \ref{fig3}.\ Note that higher repumping rates cause a linewidth broadening in the overall scale \cite{SM}.\ At such stronger $w$, for example $w$ $=$ $10\Gamma$, the excited state population $\langle\hat\sigma_\mu^{ee}\rangle_c$ $\gtrsim$ $0.9$, which even becomes totally inverted as $w$ $\gtrsim$ $50\Gamma$.\ This saturation of atomic excitations indicates that the linewidth broadening results entirely from the incoherent repumping fields.\

\begin{figure}[t]
\centering
\includegraphics[width=8.0cm,height=4.5cm]{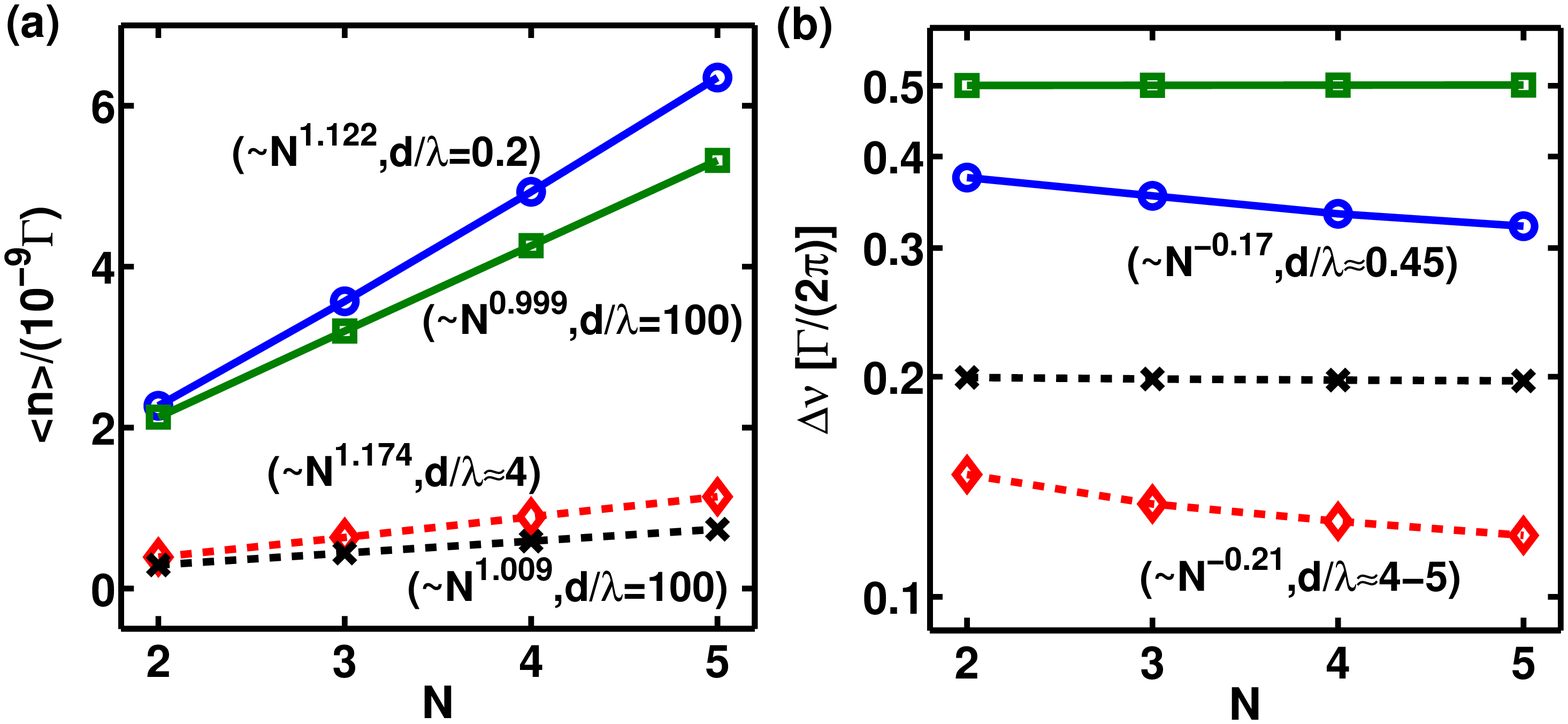}
\caption{(Color online) Scaling of cavity field and its spectral linewidth as a number of an equidistant atomic array.\ The photon number (a) and associated linewidth (b) in logarithmic scale are plotted for $N$ $=$ $2-5$ with $w$ $=$ $0.1$ (dash-$\times$ and $\Diamond$) and $2\Gamma$ (solid-$\square$ and $\bigcirc$).\ The scalings are fitted for the maximal photon number and the smallest linewidth, where various optimal $\xi$'s are denoted in parentheses.\ Almost noninteracting regimes at $\xi$ $=$ $100\lambda$ [$\times$ and $\square$ for both (a) and (b)] are included to compare the scalings with a finite LRDDI, which are nearly independent of $N$.}\label{fig4}
\end{figure}

{\it Scaling of $N$.} A natural next question is to ask what would the number of particles do for a SL.\ To get an intuitive perspective on this effect, we make a discrete increase of $N$, and limit ourselves in the setting of an equidistant atomic array.\ In Fig. \ref{fig4} we plot the maximal cavity fields of Fig. \ref{fig2} and the minimal linewidths of Fig. \ref{fig3} respectively for both weak and strong repumping rates.\ In subplot (a), the scaling of the photon number in a noninteracting regime goes linear as expected.\ On the contrary at some finite $\xi$'s with a significant or moderate LRDDI for a strong or weak $w$ respectively, the maximum of the cavity field goes up in a scaling of $10\%$ more than the linear dependence.\ Approximately ten times enhancement of the photon number comparing the one in a nonnteracting regime can be derived when $N$ $=$ $10^6$ for a weak $w$.\ This enhancement also reflects on the reduction of linewidth in subplot (b).\ Using the same projection of $N$, the linewidth is reduced to $0.01\Gamma$ ($0.04\Gamma$) by a factor of $20$ ($12.5$) compared to the noninteracting value for a weak (strong) $w$.\ These prominent scaling factors for the linewidths indicate a significant improvement on the performance of the steady state SL utilizing LRDDI.

The results and the properties derived here are all scaled by the effective spontaneous emission rate $\Gamma$.\ This $\Gamma$ however varies in different experimental setups.\ Here we consider two main atomic systems that are presently operated to realize the steady state SL.\ One is for the dipole-forbidden transitions of $^3P_0$-$^1S_0$ and $^3P_1$-$^1S_0$ in alkaline-earth-metal isotopes $^{87}$Sr and $^{88}$Sr respectively.\ The effective relaxation rates are taken as $1$ Hz \cite{Meiser2009} and $7.5$ kHz \cite{Norcia2016} with lasing wavelengths of $698$ and $689$ nm respectively, where the former has included inhomogeneous processes other than the intrinsic intercombination rate $\sim$ $0.01$ Hz.\ The other atomic setup is $^{87}$Rb atoms using a Raman lasing transition with a tunable $\Gamma$ $\approx$ $2-60$ Hz \cite{Bohnet2012}.\ Taking advantage of LRDDI in a SL with the optimal atomic separations, we have extrapolated as small as a reduction of its linewidth by two order of magnitude, reaching a level of $10$ mHz if $\Gamma$ $=$ $1$ Hz and $N$ $=$ $10^6$.\ The optimal inter-particle distance of $\xi$ $=$ $4$ and $0.45$ suggests an ensemble density of $3.1\times 10^{10}$ cm$^{-3}$ and $2.2\times 10^{13}$ cm$^{-3}$ respectively in terms of $D_1$ transition wavelength ($795$ nm) of $^{87}$Rb, which are typical in present cold atom experiments.

\begin{figure}[t]
\centering
\includegraphics[width=8.0cm,height=4.5cm]{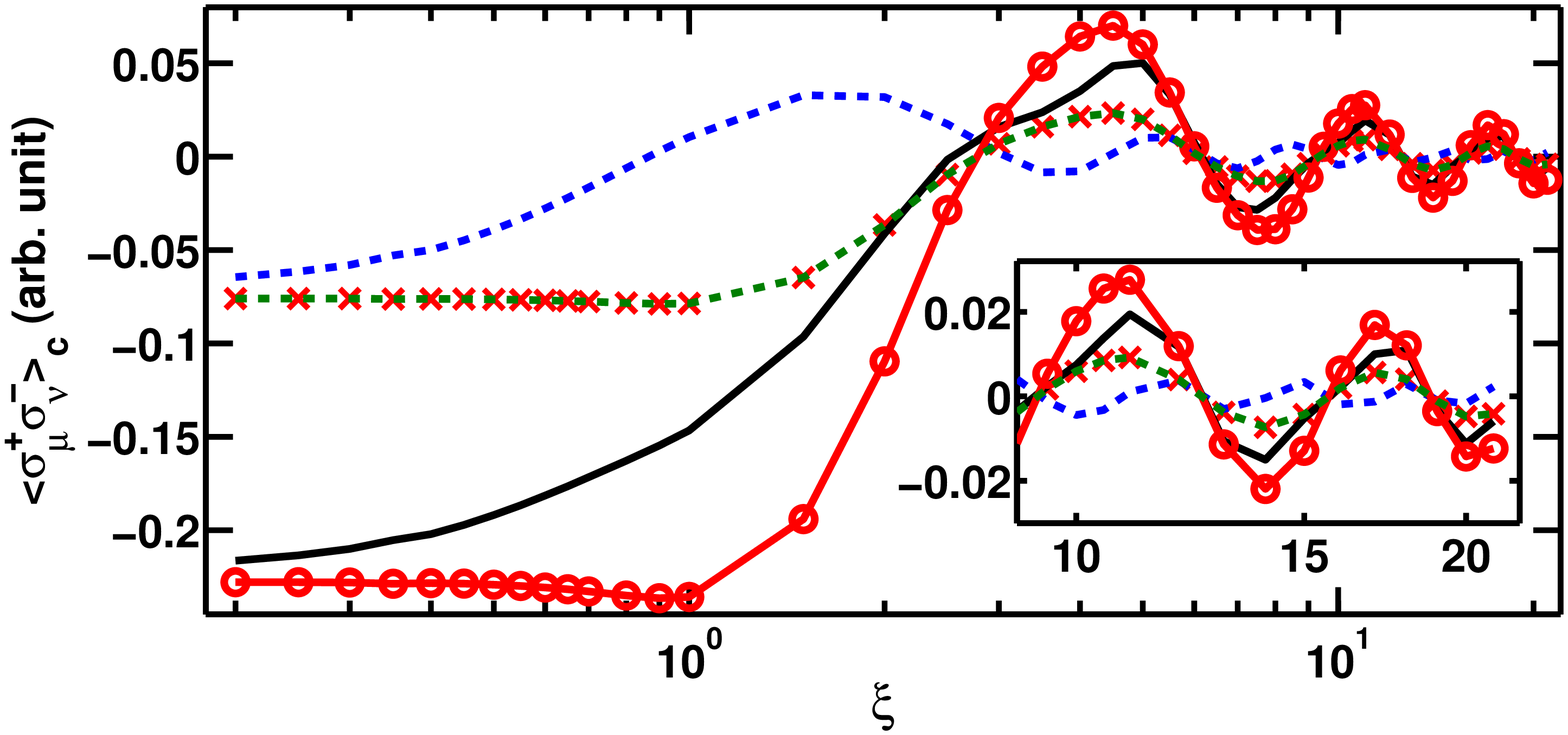}
\caption{(Color online) Atom-atom coherences for a SL of $N$ $=$ $3$.\ Three atom-atom coherences of $\langle\hat\sigma_1^+\hat\sigma_2^-\rangle_c$ on top of $\langle\hat\sigma_2^+\hat\sigma_3^-\rangle_c$ (dash-$\times$), and $\langle\hat\sigma_1^+\hat\sigma_3^-\rangle_c$ (dash) are plotted.\ The sum of all three coherences (solid) is compared to $3\langle\hat\sigma_1^+\hat\sigma_2^-\rangle_c$ ($\bigcirc$), showing an obvious deviation even in the limit of large $\xi$ in the inset.\ Here $w$ $=$ $0.1\Gamma$ and the other parameters are the same as in Fig. \ref{fig2}.}\label{fig5}
\end{figure}

As a final remark, we point out the ambiguity in treating a large ensemble atomic system in terms of a part of the whole, and the complexity arising in many-body correlations.\ In Fig. \ref{fig5} we take a SL of $N$ $=$ $3$ as an example and plot three possible atom-atom coherences of the system.\ It is evident that $3\langle\hat\sigma_1^+\hat\sigma_2^-\rangle_c$ $\neq$ $\langle\hat\sigma_1^+\hat\sigma_2^-\rangle_c$ $+$ $\langle\hat\sigma_1^+\hat\sigma_3^-\rangle_c$ $+$ $\langle\hat\sigma_2^+\hat\sigma_3^-\rangle_c$ throughout the parameter regime of $\xi$ we consider.\ The deviation is obvious at small $\xi$ since $\langle\hat\sigma_1^+\hat\sigma_3^-\rangle_c$ discords with the other two coherences of equal atomic separations, while at large $\xi$, it still remains finite.\ In present cold atom experiments operating with a confined atomic system, we expect that the many-body pairwise correlations would significantly modify the steady state SL in a cavity-QED system beyond the mean-field treatment.

In conclusion, we theoretically study the effect of LRDDI on a steady state SL.\ The cavity photon number and atom-atom coherence oscillate as the inter-particle distance varies.\ The alternating sign changes in the atom-atom coherence indicate a critical transition to the enhancement or reduction of the cavity fields.\ The optimal inter-particle separation is located to allow for an even smaller spectral linewidth than the one in a negligible LRDDI regime.\ In the setting of an equidistant atomic array, we study the SL properties with a scaling of $N$.\ We demonstrate that LRDDI enables a better performance of a steady state SL with optimal operation parameters, which can facilitate advanced precision measurements and optical clocks.

{\it Acknowledgements.} This work is supported by the Ministry of Science and Technology (MOST), Taiwan, under Grant No. MOST-101-2112-M-001-021-MY3 and No. MOST-103-2112-M-001-011.\ We are also grateful for stimulating discussions with Y.-C. Chen, M.-S. Chang, and S.-Y. Lan, and the support of NCTS in Hsinchu, Taiwan.

\clearpage
\section{Supplemental materials for Superradiant laser: Effect of long-ranged dipole-dipole interaction}
\section{Heisenberg equations of motion}
From the Hamiltonian in a Lindblad form in the main paper, we proceed to write down the coupled Heisenberg equations of motion.\ We assume $\langle\hat{a}\rangle$ $=$ $\langle\hat{\sigma}_\mu^\pm\rangle$ $=$ $0$, which respectively indicates that photons are in the Fock states and atoms acquire no macroscopic dipole moment in the cavity QED system.\ The equations of motion can be derived as follows, and first we consider the excited state population,
\bea
\frac{d\hat{\sigma}_\mu^{ee}}{dt}&=&w\hat{\sigma}_\mu^{gg}-\Gamma\hat{\sigma}_\mu^{ee}+\frac{ig}{2}(\hat{a}^\dag\hat{\sigma}_\mu^- - \hat{\sigma}_\mu^+\hat{a})-\sum_{\mu'\neq\nu}^N\sum_{\nu}^N\frac{F_{\mu'\nu}}{2}\nonumber\\
&\times&\left(\hat{\sigma}_{\mu'}^+\hat{\sigma}_\nu^-\hat{\sigma}_\mu^{ee} + \hat{\sigma}_\mu^{ee}\hat{\sigma}_{\mu'}^+\hat{\sigma}_\nu^- - 2\hat{\sigma}_{\mu'}^+\hat{\sigma}_\mu^{ee}\hat{\sigma}_\nu^-\right)\nonumber\\
&-&\sum_{\mu'\neq\nu}^N\sum_{\nu}^N iG_{\mu\nu}\left(\hat{\sigma}_\mu^{ee}\hat{\sigma}_{\mu'}^+\hat{\sigma}_\nu^- - \hat{\sigma}_{\mu'}^+\hat{\sigma}_\nu^-\hat{\sigma}_\mu^{ee}\right),\label{excited}
\eea
where the excited state population is repumped from the ground state population, and the cavity field builds up the atom-field correlation as expected.\ In addition we see an extra dynamical coupling from atom-atom and excited state correlations due to the induced dipole-dipole interaction.\ Here we introduce the generalized cumulant expansion \cite{Kubo1962} which allows to neglect small higher order correlations and keeps up to second-order correlations of atoms or photons respectively.\ Essentially we expand the mean values of the quantum operators in terms of lower order cumulants of correlations.\ For example $\langle\hat{X}_1\hat{X}_2\rangle_c$ = $\langle\hat{X}_1\hat{X}_2\rangle$ - $\langle\hat{X}_1\rangle\langle\hat{X}_2\rangle$ for two quantum operators $\hat{X}_{1,2}$.

Applying the cumulant expansion to Eq. (\ref{excited}) where we note that $\langle\hat{\sigma}_\mu^{ee}\rangle$ $=$ $\langle\hat{\sigma}_\mu^{ee}\rangle_c$ and $\langle\hat{a}^\dag\hat{\sigma}_\mu^-\rangle$ $=$ $\langle\hat{a}^\dag\hat{\sigma}_\mu^-\rangle_c$, we realize that the expansion from long-ranged dipole-dipole (LRDD) interaction leave us the term like $\langle\hat{\sigma}_{\mu'}^+\hat{\sigma}_\nu^-\hat{\sigma}_\mu^{ee}\rangle_c$.\ Take a closer look at this three-operator cumulant, we cannot just throw it away as small higher order cumulants since they involve lower order cumulants intrinsically.\ So seemingly third-order correlation can be reduced to the second-order one if for example $\hat{\sigma}_\nu^-\hat{\sigma}_\mu^{ee}$ $=$ $\hat{\sigma}_\nu^-\delta_{\nu,\mu}$.\ Therefore we have from Eq. (\ref{excited}) after carefully keeping the second-order cumulants,
\bea
\frac{d\langle\hat{\sigma}_\mu^{ee}\rangle_c}{dt}&=&w\langle\hat{\sigma}_\mu^{gg}\rangle_c-\Gamma\langle\hat{\sigma}_\mu^{ee}\rangle_c
+\frac{ig}{2}(\langle\hat{a}^\dag\hat{\sigma}_\mu^-\rangle_c - \langle\hat{\sigma}_\mu^+\hat{a}\rangle_c)\nonumber\\
&-&\sum_{\nu\neq\mu}^N\frac{F_{\mu\nu}+i2G_{\mu\nu}}{2}\langle\hat{\sigma}_\mu^+\hat{\sigma}_\nu^-\rangle_c\nonumber\\
&-&\sum_{\nu\neq\mu}^N\frac{F_{\mu\nu}-i2G_{\mu\nu}}{2}\langle\hat{\sigma}_\nu^+\hat{\sigma}_\mu^-\rangle_c,\label{excited2}
\eea
where the last two lines show the effects from LRDD, which cause collective population redistribution from all the atom-atom correlations in the ensemble.\ If we assume $F_{\mu\nu}$ $\rightarrow$ $\Gamma\delta_{\mu,\nu}$ and $G_{\mu,\nu}$ $=$ $0$, the above reduces to a single-particle fashion \cite{Meiser2009} where many-body effect plays no role.

We continue to formulate the equations of motion including the many-body effect from LRDD.\ The atom-field correlation then follows
\bea
\frac{d\langle\hat{a}^\dag\hat{\sigma}_\mu^-\rangle_c}{dt}&=&\Big(i\delta-\frac{w+\kappa+\Gamma}{2}\Big)\langle\hat{a}^\dag\hat{\sigma}_\mu^-\rangle_c+\frac{ig}{2}\nonumber\\
&\times&\Big[\langle\hat{a}^\dag\hat{a}\rangle_c\langle2\hat{\sigma}_\mu^{ee}-1\rangle_c+\langle\hat{\sigma}_\mu^{ee}\rangle_c+\sum_{\nu\neq\mu}^N\langle\hat{\sigma}_\nu^+\hat{\sigma}_\mu^-\rangle_c\Big]\nonumber\\
&-&\sum_{\nu\neq\mu}^N\frac{F_{\mu\nu}+i2G_{\mu\nu}}{2}\langle1-2\hat{\sigma}_\mu^{ee}\rangle_c\langle\hat{a}^\dag\hat{\sigma}_\nu^-\rangle_c,
\eea
where the detuning is $\delta$ $=$ $\omega_c-\omega_a$, and we have neglected the third order cumulants  $\langle\hat{a}^\dag\hat{a}\hat{\sigma}_\mu^{ee}\rangle_c$ and $\langle\hat{\sigma}_\mu^{ee}\hat{a}^\dag\hat{\sigma}_\nu^-\rangle_c$.\ The above first two lines couple atom-field correlation to the population operator and atom-atom correlation, while the third line dynamically couples all the atoms in the ensemble due to the LRDD interaction.\ We note that extra correlation arises between the excited state population and atom-field correlation operators in the third line, which is a third order correlation which we have neglected.

The atom-atom correlation evolves as
\bea
\frac{d\langle\hat{\sigma}_\mu^+\hat{\sigma}_\nu^-\rangle_c}{dt}&=&-(w+\Gamma)\langle\hat{\sigma}_\mu^+\hat{\sigma}_\nu^-\rangle_c +\frac{g}{2i}\nonumber\\
&\times&\left[\langle\hat{a}^\dag\hat{\sigma}_\nu^-\rangle_c \langle2\hat{\sigma}_\mu^{ee}-1\rangle_c-\langle\hat{\sigma}_\mu^+\hat{a}\rangle_c \langle2\hat{\sigma}_\nu^{ee}-1\rangle_c\right]\nonumber\\
&-&\sum_{m\neq\nu}^N\frac{F_{m\nu}+i2G_{m\nu}}{2}\langle\hat{\sigma}_\mu^+\hat{\sigma}_m^-\rangle_c\langle 1-2\hat{\sigma}_\nu^{ee}\rangle_c\nonumber\\
&-&\sum_{m\neq\mu}^N\frac{F_{\mu m}-i2G_{\mu m}}{2}\langle\hat{\sigma}_m^+\hat{\sigma}_\nu^-\rangle_c\langle 1-2\hat{\sigma}_\mu^{ee}\rangle_c,
\eea
where we neglect higher order cumulants of $\langle\hat{\sigma}_m^+\hat{\sigma}_\nu^-\hat{\sigma}_{\mu\neq\nu}^{ee}\rangle_c$ and the others involving four operators.\ The last two lines in the above indicate collective decay channels for atom-atom correlations again due to the LRDD interaction.\ Whether the correlations are enhanced or reduced depend on the signs of other pairwise atom-atom correlations and excited state populations.\ We note that when $m$ $=$ $\mu$ in the third line for example, the cumulant becomes $\sim$ $\langle\hat{\sigma}_\mu^{ee}\rangle_c\langle\hat{\sigma}_\nu^{ee}\rangle_c$ which denotes a coupling between the atom-atom and density-density correlations via collective decay channels.  

Finally we have the equation of motion for the photon number,
\bea
\frac{d\langle\hat{a}^\dag\hat{a}\rangle_c}{dt}=-\kappa\langle\hat{a}^\dag\hat{a}\rangle_c+\frac{g}{2i}\sum_{\mu=1}^N(\langle\hat{a}^\dag\hat{\sigma}_\mu^-\rangle_c - \langle\hat{\sigma}_\mu^+\hat{a}\rangle_c).\label{photon}
\eea
From Eqs. (\ref{excited2})-(\ref{photon}), the closed coupled equations are formed, where we can solve for the steady state solutions numerically.

\section{Cavity field and its linewidth}
\begin{figure}[b]
\centering
\includegraphics[width=8.0cm,height=4.2cm]{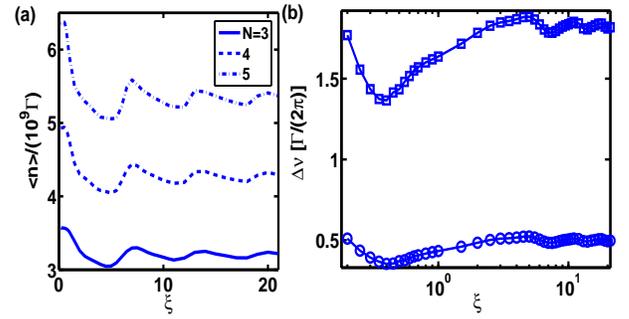}
\caption{(Color online) Cavity fields for SL of $N$ $=$ $3-5$ and linewidths for $N$ $=$ $3$ at strong repumping rates.\ In bad-cavity limit we choose the coupling constant $g$ $=$ $40\Gamma$ and cavity loss rate $\kappa$ $=$ $10^6\Gamma$.\ (a) The cavity photon number has a similar oscillatory as $N$ $=$ $2$ for a repumping rate $w$ $=$ $2\Gamma$.\ (b) The linewidths also have similar patterns as $N$ $=$ $2$ for strong repumping rates $w$ $=$ $2$ ($\bigcirc$) and $10\Gamma$ ($\square$).}\label{s1}
\end{figure}

Here we show the results of cavity photon number for $N$ $=$ $3-5$, and the linewidth for SL of $N$ $=$ $3$ at two different repumping rates in Fig. \ref{s1}.\ The cavity fields have the similar dependences on $\xi$ in (a).\ The linewidth also has a similar pattern for strong repumping rates but the overall scale for stronger repumping rate is larger.

\end{document}